\documentclass[preprint,12pt]{elsarticle}

\usepackage[margin=0.80in]{geometry}
\usepackage{float}
\usepackage{adjustbox}
\usepackage{wrapfig}
\usepackage{soul}
\usepackage{footnote}
\usepackage{lineno}
\usepackage[utf8x]{inputenc} 

\usepackage[bookmarks=false]{hyperref}
\usepackage{listings}

\lstdefinestyle{R}{
    language        = R,
    frame           = lines, 
    basicstyle      = \footnotesize,
    keywordstyle    = \color{blue},
    stringstyle     = \color{green},
    commentstyle    = \color{red}\ttfamily
}

\makeatletter
\def\ps@pprintTitle{%
 \let\@oddhead\@empty
 \let\@evenhead\@empty
 \def\@oddfoot{\centerline{\thepage}}%
 \let\@evenfoot\@oddfoot}
\makeatother

\usepackage{multirow}
\usepackage{graphics}
\usepackage[dvips]{color}
\usepackage{graphicx}

\usepackage{caption}
\usepackage{subcaption}

\usepackage{amssymb,amstext,amsmath}
\usepackage{float}
\usepackage{booktabs}
\usepackage{array,ragged2e}
\usepackage[table]{xcolor}
\usepackage{multirow}
\usepackage{algorithm}
\usepackage{algpseudocode}
\usepackage{pifont}

 \biboptions{sort&compress}

\begin{document}
\begin{frontmatter}

\title{Bayesian Parameter Estimations for Grey System Models in Online Traffic Speed Predictions}

\author[MCSaddress,rvt2]{Gurcan Comert\corref{mycorrespondingauthor}}
\cortext[mycorrespondingauthor]{Corresponding author}
\ead{gurcan.comert@benedict.edu}

\author[MCSaddress]{Negash Begashaw}
\ead{negash.begashaw@benedict.edu}

\author[PEaddress]{Negash G. Medhin}
\ead{ngmedhin@ncsu.edu}

\address[MCSaddress]{Department of Computer Science, Physics, and Engineering, Benedict College, Columbia, SC 29204 USA}
\address[rvt2]{Information Trust Institute, University of Illinois Urbana-Champaign, Urbana, IL 61801 USA}
\address[PEaddress]{Department of Mathematics, North Carolina State University, Raleigh, NC 27695 USA}

\begin{abstract}

This paper presents Bayesian parameter estimation for first order Grey system models’ parameters (or sometimes referred to as hyperparameters). There are different forms of first-order Grey System Models. These include $GM(1,1)$, $GM(1,1| \cos(\omega t)$, $GM(1,1| \sin(\omega t)$, and $GM(1,1| \cos(\omega t), \sin(\omega t)$. The whitenization equation of these models is a first-order linear differential equation of the form
\[
\frac{dx}{dt} + a x = f(t)
\]
where $a$ is a parameter and $f(t) = b$  in $GM(1,1|)$ , $f(t) = b_1\cos(\omega t) + b_2$ in $GM(1,1| cos(\omega t)$, $f(t) = b_1\sin(\omega t)+b_2$ in
$GM(1,1| \sin(\omega t)$, $f(t) = b_1\sin(\omega t) + b_2\cos(\omega t) + b_3$ in $GM(1,1| \cos(\omega t), \sin(\omega t)$,
$f(t) = b x^2$ in Grey Verhulst model (GVM),
 and where $b, b_1, b_2, b_3$ are parameters.  The results from Bayesian estimations are compared to the least square estimated models with fixed $\omega$.  We found that using rolling Bayesian estimations for GM parameters can allow us to estimate the parameters in all possible forms.  Based on the data used for the comparison, the numerical  results showed that models with Bayesian parameter estimations are up to 45\% more accurate in mean squared errors.

\end{abstract}

\end{frontmatter}

\section{Introduction}
Grey systems models (GMs) first introduced by Professor Julong in 1989 \cite{julong1989introduction}, have been applied to many areas of interest and are proved to be good prediction models in a wide range of applications especially in finance. 
In GM, we first estimate parameters using the method of least squares then we plug in the estimated parameters in the solution of the first order linear differential equation used as the whitenization of the grey model to make predictions. For real time usage, a window size of 4 previous data points is found to be effective for updating the parameters \cite{kayacan2010grey,bezuglov2016short,comert2020gaussian,COMERT2021114972,comert2021grey}. WinBUGS models use $R^2$ but we do not use $R^2$  in this paper.
In a previous work, we used GMs for short-term speed and travel time predictions in transportation \cite{bezuglov2016short}. In \cite{COMERT2021114972}, improved grey system models were used to predict short-term traffic parameters (e.g., speed, travel time, flow, and occupancy). \cite{comert2021grey} showed the effectiveness of GMs in short-term traffic signal queue lengths which are highly nonlinear (cyclic). In \cite{comert2020gaussian}, the authors compared the performance of GMs in short-term and long-term (aggregated) traffic parameter predictions. The authors in \cite{comert2020evaluating} used GMs for air quality index predictions. 

In the literature, there are only two previous papers \cite{hsu2007forecasting,sinha2020forecasting} that dealt with Bayesian treatment of grey system models. The papers in \cite{hsu2007forecasting,sinha2020forecasting}] dealt with Bayesian treatment for only GM(1,1) models. It is observed that there is a limited improvement in GM(1,1). In our study in this paper, we will show the effectiveness of the online Bayesian approach (utilizing WinBUGS and R) in the grey models GM(1,1), GM Verhulst, and GM-trigonometric models using classical volatile time series traffic speed data that contains more than 300 data values. For our work in this paper, we used a sample of 20 annual values for forecasting.

\section{Overview of the Grey Models Used}
This section brief introduces of the Grey system models where we applied Bayesian parameter estimation to approximate the Grey parameters in online traffic speed prediction.
\newline
The series $X^{(0)}=(x^{(0)}(1),x^{(0)}(2),...,x^{(0)}(n))$ represent positive observations of a process and $X^{(1)}=(x^{(1)}(1),x^{(1)}(2),...,x^{(1)}(n))$ is an accumulation sequence of $X^{(0)}$ where 
\begin{equation}
x^{(1)}(k) = \sum_{i=1}^{k}{x^{(0)}(i)}
\label{eq:accumulated_seq}
\end{equation}

The sequence $Z^{(1)}=(z^{(1)}(2),z^{(1)}(3),...,z^{(1)}(n))$ can be taught as a weighted mean sequence of $X^{(1)}$ where
\begin{equation}
z^{(1)}(k) = \frac{z^{(1)}(k-1)+z^{(1)}(k)}{2}, \forall k = 2,3,\cdots,n
\label{eq:Z_1}
\end{equation}

The following equation gives the basic form of GM\_11 (GM(1,1)) with parameters $a$-coefficient and $b$-intercept.
\begin{equation}
x^{(0)}(k) + az^{(1)}(k) = b
\label{eq:basic_GM}
\end{equation}

If $(\hat{a},\hat{b})^T=(a,b)^T$ and
\begin{eqnarray}
Y = \left[
\begin{array}{c}
x^{(0)}(2) \\
x^{(0)}(3) \\
\vdots \\
x^{(0)}(n) \\
\end{array}
\right],
B = \left[
\begin{array}{cc}
-z^{(1)}(2) & 1 \\
-z^{(1)}(3) & 1 \\
\vdots & \vdots \\
-z^{(1)}(n) & 1 \\
\end{array}
\right].
\label{eq:YandB}
\end{eqnarray}

then, as in~\cite{liu2006grey}, the least-squares estimate of the GM\_11 model is~$(\hat{a},\hat{b})^T=(B^TB)^{-1}B^TY$. $\hat{x}^{(0)}(k)$ and $\hat{x}^{(1)}(k)$ are the predicted sequence and the accumulated time response sequence of GM\_11 at time $k$ respectively. The following prediction equation is used:

\begin{equation}
\hat{x}^{(0)}(k+1)=\left(1-e^a\right)\left(x^{(0)}(1)-\frac{b}{a}\right)e^{-ak}, k=1,2,...,n
\label{eq:model_x_0_solution}
\end{equation}

Eq.~(\ref{eq:model_x_0_solution}) is the prediction equation that outputs values $\forall k=2,3,...,n$. In the rolling window framework: $x^{(0)}(k+1),x^{(0)}(k+2),...,x^{(0)}(k+w)$, where $w \geq 4$ is the window size ( $w=4$ is found to produce very good results \cite{bezuglov2016short}). 

For our work in this paper, we used the Grey Verhulst (GVM) model, Grey models with cosine, sine, and a linear combination of sine and cosine terms. In what follows we describe these grey system models in detail.

\subsection{The Grey Verhulst model (GVM)}
\label{sctGVM}
The Grey Verhulst model (GVM) is utilized for data behaving more nonlinearly \cite{liu2010grey}. The underlying structure of the GVM is given by the following equation \cite{kayacan2010grey,bezuglov2016short}.
%
%
\begin{equation}
x^{(0)}(k)+az^{(1)}(k)=b\left(z^{(1)}(k)\right)^2
\label{eq:verhulst_model}
\end{equation}

The whitenization equation of GVM is:
\begin{equation}
\frac{dx^{(1)}}{dt} + ax^{(1)} = b\left(x^{(1)}\right)^2
\label{eq:verhulst_whitenization}
\end{equation}

Like in GM(1,1), the least squares estimate is applied to find~$(\hat{a},\hat{b})^T=(B^TB)^{-1}B^TY$, where
\begin{eqnarray}
Y = \left[
\begin{array}{c}
x^{(0)}(2) \\
x^{(0)}(3) \\
\vdots \\
x^{(0)}(n) \\
\end{array}
\right],
B = \left[
\begin{array}{cc}
-z^{(1)}(2) & z^{(1)}(2)^2 \\
-z^{(1)}(3) & z^{(1)}(3)^2 \\
\vdots & \vdots \\
-z^{(1)}(n) & z^{(1)}(n)^2 \\
\end{array}
\right].
\label{eq:YandB_VerhulstModel}
\end{eqnarray}

The predictions $\hat{x}^{(0)}(k+1)$ are calculated using Eq.~(\ref{eq:verhulst_x_0_solution}).
\begin{equation}
\hat{x}^{(0)}(k+1)=[\frac{ax^{(0)}(1)\left(a-bx^{(0)}(1)\right)}{bx^{(0)}(1)+\left(a-bx^{(0)}(1)\right)e^{a(k-1)}}][\frac{\left(1-e^a\right)e^{a\left(k-2\right)}}{bx^{(0)}(1)+\left(a-bx^{(0)}(1)\right)e^{a(k-2)}}]
\label{eq:verhulst_x_0_solution}
\end{equation}
%
\vspace{-10 pt}
\subsection{GM(1,1$|sin(\omega t)$) model}
The idea in the Grey trigonometric models model is to enhance the prediction of the original GM(1,1) under seasonal behavior. This is accomplished by introducing trigonometric terms on the right-hand side of the whitenization equation. To improve the classic GMs, we simply utilized a sine function, a cosine function, a combination of sine and cosine functions, and a combination of sine, cosine, and multiplied by an exponential term. 
\newline 
\indent

GM(1,1$|sin(\omega t)$) model was used in \cite{shuhua2012city} and its whitenization equation is the first order linear differential equation described in Eq.~(\ref{eq:whitenization_5}).
\begin{equation}
\frac{dx^{(1)}}{dt} + ax^{(1)}(k) = b_1 sin(\omega t)+b_2
\label{eq:whitenization_5}
\end{equation}

With the same approach as GM(1,1) and GVM models, the model parameters $(\hat{a},\hat{b}_1,\hat{b}_2)^T=(a,b_1,b_2)^T$ are estimated by $(\hat{a},\hat{b}_1,\hat{b}_2)=(B^TB)^{-1}B^TY$ where $Y=[x^{(0)}(2),...,x^{(0)}(n)]^T$ is given above and $B$ is:
\begin{eqnarray}
B = \left[
\begin{array}{ccc}
-z^{(1)}(2) & sin(\omega 2)& 1 \\
-z^{(1)}(3) & sin(\omega 3)&1 \\
\vdots & \vdots \\
-z^{(1)}(n) & sin(\omega n)&1 \\
\end{array}
\right].
\label{eq:YandB5}
\end{eqnarray}

Terms $\hat{x}^{(0)}(k)$ and $\hat{x}^{(1)}(k)$ represent the predictions and the accumulated time response sequence of GM(1,1$|sin(\omega t)$) at time $k$ respectively. Then, the latter can be obtained by solving the following equation.
\begin{equation}\label{eq:model_5_solution}
\left.\begin{aligned}
x^{(1)}(t+1)&=(x^{(1)}(1)+\frac{b_1\omega}{a^2+\omega^2}-\frac{b}{a})e^{-at}+\\
&\frac{b_1\omega}{a^2+\omega^2}(a sin(\omega t)-\omega cos(\omega t))+\frac{b}{a}
\end{aligned}\right.
\end{equation}
Using the initial condition $x^{(1)}(1)$=$x^{(0)}(1)$ $,\forall k=2,3,...,n$, the reduced (non-cumulative or differenced) of $\hat{x}^{(0)}(k+1)$ in Eq.~(\ref{eq:model_5_solution}) can be calculated as $\hat{x}^{(1)}(k+1)-\hat{x}^{(1)}(k)$. Thus, using Eq.~(\ref{eq:model_5_solution}), the following prediction is obtained.

\begin{equation}\label{eq:model_5}
\left.\begin{aligned}
&\hat{x}^{(0)}(k+1)=(1-e^a)(x^{(0)}(1)+\frac{b_1\omega}{a^2+\omega^2}-\frac{b}{a})e^{-ak}+ \\
&\frac{b_1\omega}{a^2+\omega^2}(a(sin(\omega k)-sin(\omega (k-1)))-\\
&\omega (cos(\omega (k-1))-cos(\omega k)))
\end{aligned}\right.
\end{equation}
%
\vspace{-10 pt}
\subsection{GM(1,1$|cos(\omega t)$) model}
Following GM(1,1$|sin(\omega t)$) steps, the whitenization equation of the model with cosine function is represented by the first-order linear differential equation.
\begin{equation}
\frac{dx^{(1)}}{dt} + ax^{(1)}(k) = b_1 cos(\omega t)+b_2
\label{eq:whitenization_6}
\end{equation}

If $(\hat{a},\hat{b}_1,\hat{b}_2)^T=(a,b_1,b_2)^T=(B^TB)^{-1}B^TY$ and
\begin{eqnarray}
B = \left[
\begin{array}{ccc}
-z^{(1)}(2) & cos(\omega 2)& 1 \\
-z^{(1)}(3) & cos(\omega 3)&1 \\
\vdots & \vdots \\
-z^{(1)}(n) & cos(\omega n)&1 \\
\end{array}
\right].
\label{eq:YandB6}
\end{eqnarray}

The solution of the differential equation in Eq.~(\ref{eq:whitenization_6}) is given by Eq.~(\ref{eq:model_6_solution}).
\begin{equation}\label{eq:model_6_solution}
\left.\begin{aligned}
&x^{(1)}(t+1)=Ce^{-t}+\\
&(a^{2}b_2 + b_2\omega^2 + a^2 b_1cos(\omega t) + a b_1 \omega sin(\omega t))/(a(a^2 + \omega^2))
\end{aligned}\right.
\end{equation}

where $C$ is obtained from the initial condition  $x^{(1)}(1)$=$x^{(0)}(1)$ and is given by
\begin{equation}\label{eq:model_6_C}
\left.\begin{aligned}
&C=e^{a}[x^{(0)}(1)-(a^{2}b_2 + b_2\omega^2 + \\
&a^2 b_1cos(\omega) + a b_1 \omega sin(\omega))/(a(a^2 + \omega^2))]
\end{aligned}\right.
\end{equation}

Following the same procedure to derive Eq.~(\ref{eq:model_5}) above, using Eq.~(\ref{eq:model_6_solution}), the following prediction equation  is obtained. %
\begin{equation}\label{eq:model_6}
\left.\begin{aligned}
&\hat{x}^{(0)}(k+1)=Ce^{-ak}(1-e^{a})+ \\
&\frac{1}{a^3+a\omega^2}[-a b_1\omega(sin(\omega k)-sin(\omega (k-1)))+\\
&a^2 b_1 (cos(\omega (k-1))-cos(\omega k))]
\end{aligned}\right.
\end{equation}
\vspace{-2 pt}
\subsection{GM(1,1$|sin(\omega t),cos(\omega t)$) model}
The whitenization equation of the model with a linear combination of Sine and Cosine functions is given by the following differential equation. ~Eq.~(\ref{eq:whitenization_7}).
\begin{equation}
\frac{dx^{(1)}}{dt} + ax^{(1)}(k) = b_1 sin(\omega t)+b_2 cos(\omega t)+b_3
\label{eq:whitenization_7}
\end{equation}

If $(\hat{a},\hat{b}_1,\hat{b}_2,\hat{b}_3)^T=(a,b_1,b_2,b_3)^T=(B^TB)^{-1}B^TY$ and
\begin{eqnarray}
B = \left[
\begin{array}{cccc}
-z^{(1)}(2) &sin(\omega 2) &cos(\omega 2)& 1 \\
-z^{(1)}(3) & sin(\omega 3)&cos(\omega 3)&1 \\
\vdots & \vdots \\
-z^{(1)}(n) & sin(\omega n)&cos(\omega n)&1 \\
\end{array}
\right].
\label{eq:YandB7}
\end{eqnarray}

The solution of the differential equation in Eq.~(\ref{eq:whitenization_7}) is given by,
\begin{equation}\label{eq:model_7_solution}
\left.\begin{aligned}
&x^{(1)}(t+1)=Ce^{-t}+\\
&\frac{(\frac{b_3}{a}+cos(\omega t)(a b_2-b_1\omega)+(ab_1+b_2\omega)sin(\omega t))}{a^2+\omega^2}
\end{aligned}\right.
\end{equation}

where $C$ is obtained from the initial condition $x^{(1)}(1)$=$x^{(0)}(1)$ and is given by
\begin{equation}\label{eq:model_7_C}
\left.\begin{aligned}
&C=e^{a}[x^{(0)}(1)-\\
&\frac{(\frac{b_3}{a}+cos(\omega)(a b_2-b_1\omega)+(ab_1+b_2\omega)sin(\omega))}{a^2+\omega^2}]+\\
\end{aligned}\right.
\end{equation}

Similar to Eq.~(\ref{eq:model_5}) above, Eq.~(\ref{eq:model_7_solution}) can reduce to Eq.~(\ref{eq:model_7}).
\begin{equation}\label{eq:model_7}
\left.\begin{aligned}
&\hat{x}^{(0)}(k+1)=Ce^{-ak}(1-e^a)+\\
&\frac{1}{a^2+\omega^2}(cos(\omega k)-cos(\omega (k-1)))(a b_2-b_1\omega)+\\
&(ab_1+b_2\omega)(sin(\omega k)-sin(\omega (k-1))
\end{aligned}\right.
\end{equation}

\begin{table}[h!]
	\centering	
	\caption{\textcolor{blue}{Example Bayesian estimation results in WinBUGS for GM\_Cos Model}}
	\label{tab_1}       
	\scalebox{0.75}{
		\begin{tabular}{c c c c c c c c c c c}
			
			\hline
	&node	 &mean	 &sd	&$2.5\%$	&$25\%$	&$50\%$ &$75\%$ &$97.5\%$	&sample \\
			\hline
&alpha  &23.24886 &67.85718 &-109.9875 &-20.915  &24.230  &67.5325  &163.9750
&5000\\    
&beta1 &-0.002434616 &0.04929079 &-0.083455  &-0.018485 &-0.00174  &0.01573250  &0.08839
&5000\\    
&beta2  &30.8593 &68.14664 &-109.775 &-13.6500  &30.08  &76.030  &166.5625
&5000\\    
&omega  &0.00007833995 &0.001227439 &0.000   &0.000  &3.3085e-37  &1.1675e-24  &1.61975e-06
&5000\\      
&tau  &0.1441969 &0.1363389  &0.004042739   &0.042474978  &0.1052  &0.205025  &4.962366e-01
&5000\\ 
&deviance  &21.08159 &4.165064 &16.49  &18.080  &19.915  &22.79749  &32.61099\\	
			\noalign{\smallskip}\hline\noalign{\smallskip}
		\end{tabular}		
	}
\end{table}
\vspace{-10 pt}
\begin{table}[ht]
\centering
\caption{\textcolor{blue}{Comparison of Grey System models with least squares and Bayesian estimations}}
\label{tab_accuracy}
\scalebox{0.8}{
\begin{tabular}{r|rrrrr|rrrrr}
  \hline
 &\multicolumn{5}{c}{Least Squares Estimation} 		&		\multicolumn{5}{c}{Bayesian Parameter Estimation} \\			
 & GM11 & GVM & GM\_Sin & GM\_Cos & GM\_SinCos & GM11 & GVM & GM\_Sin & GM\_Cos & GM\_SinCos \\ 
  \hline\noalign{\smallskip}
MSE-1 & 48.50 & 23.76 & 46.37 & 29.79 & 46.37 & 48.72 & 22.69 & 45.75 & 24.48 & 29.16 \\ 
  MSE-2 & 92.59 & 36.62 & 89.84 & 58.97 & 100.41 & 90.90 & 35.79 & 86.61 & 49.39 & 55.15 \\ 
  \% imp-1 &  &  &  &  &  & -0.46 & 4.48 & 4.56 & 17.82 & 37.12 \\ 
  \% imp-2 &  &  &  &  &  & 1.82 & 2.25 & 3.59 & 16.25 & 45.07 \\ 
   \hline
\end{tabular}
}
\end{table}
   	\begin{figure}[!h]
       \centering
       \captionsetup{aboveskip=-20pt,belowskip=-5pt}
       \includegraphics[width=0.5\linewidth]{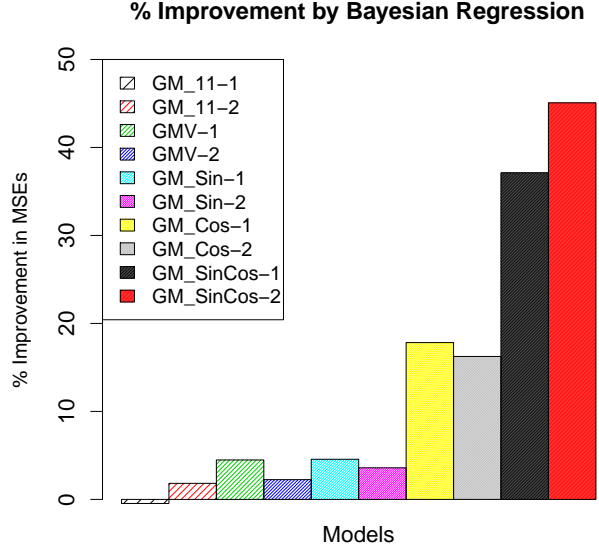}
       \caption{\textcolor{blue}{Improvements in prediction accuracy with Bayesian parameter estimation for both days}}
       \label{fig_imp}
       \end{figure}  
\vspace{-5 pt}           
\section{Computational Results} 
This paper shows real-time estimations 
 of the GM parameters using Bayesian Regression for five GMs using WinBUGS (see Example Model 1) from R in \cite{sturtz2005r2winbugs}. 
 

Priors for each model are described below.

\vspace{-20 pt}
	\begin{figure}[h]
    \centering
    \includegraphics[width=0.999\linewidth]{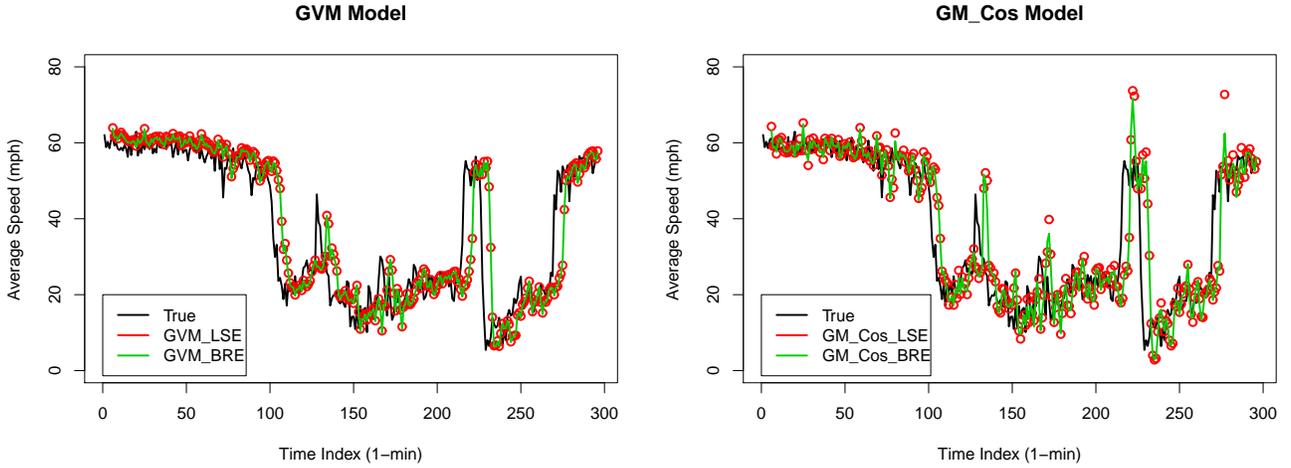}
    \caption{\textcolor{blue}{Example one step speed predictions on February 18 (1)}}
    \label{fig_fq1}
    \end{figure}
           \begin{figure}[h]
            \centering
            \includegraphics[width=0.999\linewidth]{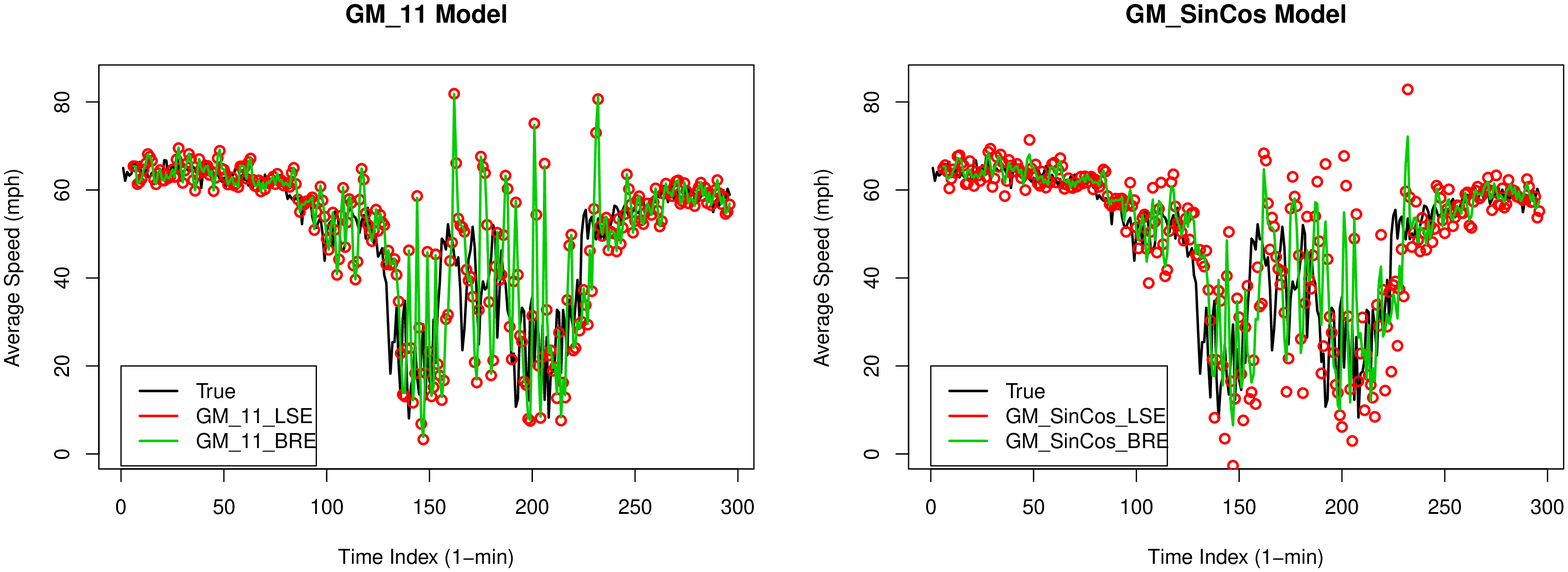}
            \caption{\textcolor{black}{Example one step speed predictions on February 19 (2)}}
            \label{fig_fq2}
            \end{figure}
\vspace{0.1in}
\begin{enumerate}
\item For GM\_11 model :  $y = a z +b$ $x^{(0)}(k)=y\sim\mathcal{N}(\mu_i,\tau)$ and coefficient of $z$ $a (\beta)\sim\mathcal{N}(0,0.001)$, intercept $b (\alpha)\sim\mathcal{N}(0,0.001)$, and $\sigma=1/\sqrt{\tau}$. 
\item For Grey Verhulst model (GVM): $y=az+bz^2$, GM\_Cos: $y = a z + b_1 cos(\omega t)+b_2$, GM\_Sin: $y = a z + b_1 sin(\omega t)+b_2$, and GM\_SinCos: $y = a z + b_1 sin(\omega t)+b_2 cos(\omega t)+b_3$,  $\omega\sim \mathcal{X}^2(0.001)$, and other parameters are initialized $\sim\mathcal{N}(0,0.001)$.  
\end{enumerate}
	\begin{figure}[!h]
    \centering
    \includegraphics[width=0.85\linewidth]{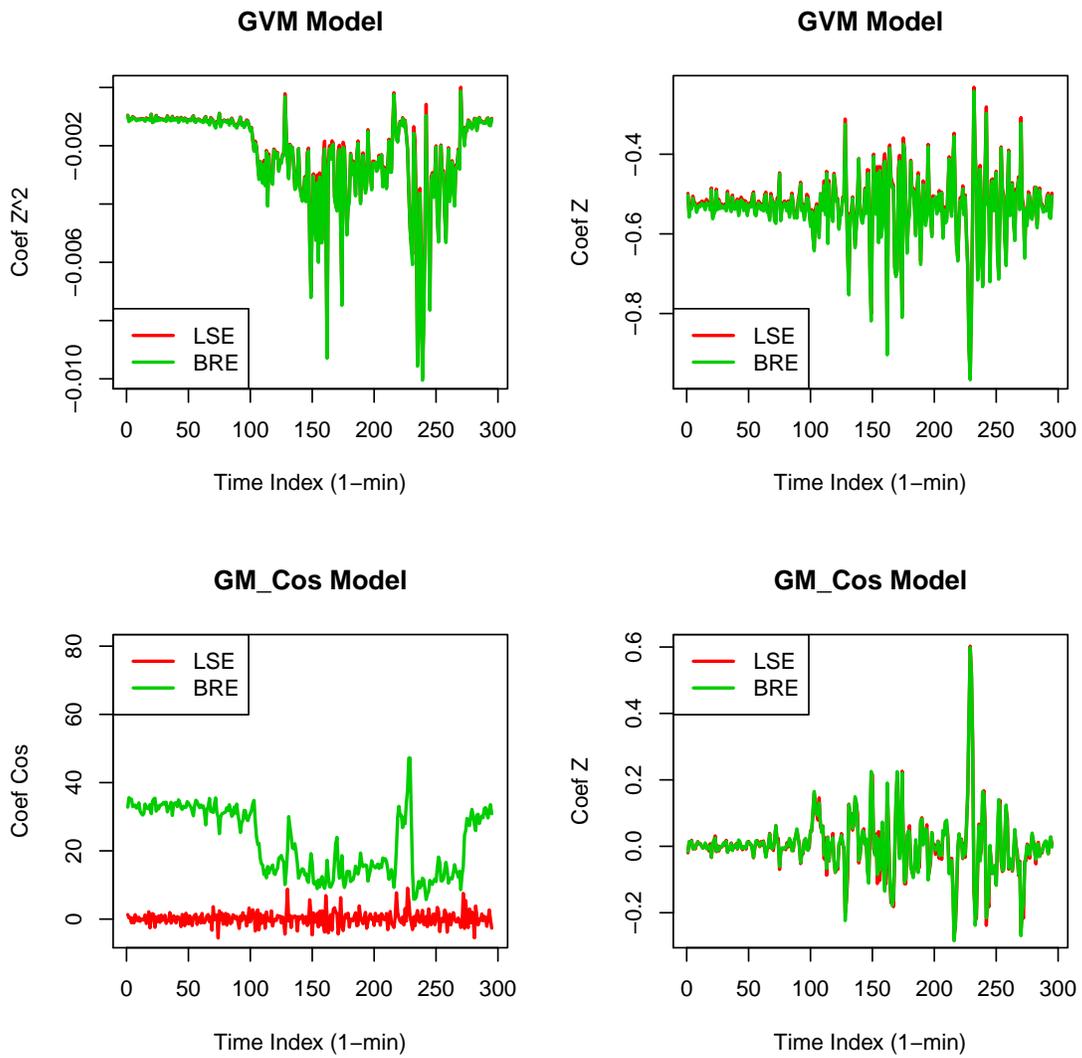}
    \caption{\textcolor{blue}{Example of evolution of parameter estimations LSE and BRE on February 18 (1)}}
    \label{fig_fqp1}
    \end{figure}
    
Note that this approach would be solely done in R using other packages or writing Gibbs samplers. One example
that uses $BayesLearn$ package $blinreg$ regression function is given in \cite{albert2009bayesian}.
It runs with similar computational times as least-squares estimators. In our work in this paper,  WinBUGS examples in \cite{spiegelhalter2003winbugs}, MCMC in \cite{albert2009bayesian}, and examples from \cite{vidakovic2017engineering} are used. Solutions and GM codes are our original codes from \cite{bezuglov2016short,COMERT2021114972} which are revised for Bayesian regression-based parameter estimations for every new data observed.
\subsection{Data Description} 
The dataset used consists of 1-minute average traffic speeds in miles per hour (mph) from loop detectors of the California PATH (Partners for Advanced Transit and Highways) program on I-$880$ for the Freeway Service Patrol Project. In this project, two days February 18 (1-Fig.~\ref{fig_fq1}) and 19 (2-Fig.~\ref{fig_fq2}) with 300 speed values are utilized. These days contain incidents that closed multiple lanes which significantly impact speed and challenge prediction models \cite{comert2013online,comert2016adaptive}.

In this paper, instead of least squares, we use Bayesian regression estimation for the parameters $(a,b_1, b_2, b_3$, and $\omega)$ using the data in the matrix in Eq.~(\ref{eq:YandB}). The results of Bayesian regression are given in Table~\ref{tab_1} for the GM Cosine model for one window ($4$ speed values). Note that we adopted a rolling horizon framework to update the parameters with every new observation and use three previous speed values. The approach is adaptive and similar to filtering.
       	\begin{figure}[!h]
            \centering
            \includegraphics[width=0.8\linewidth]{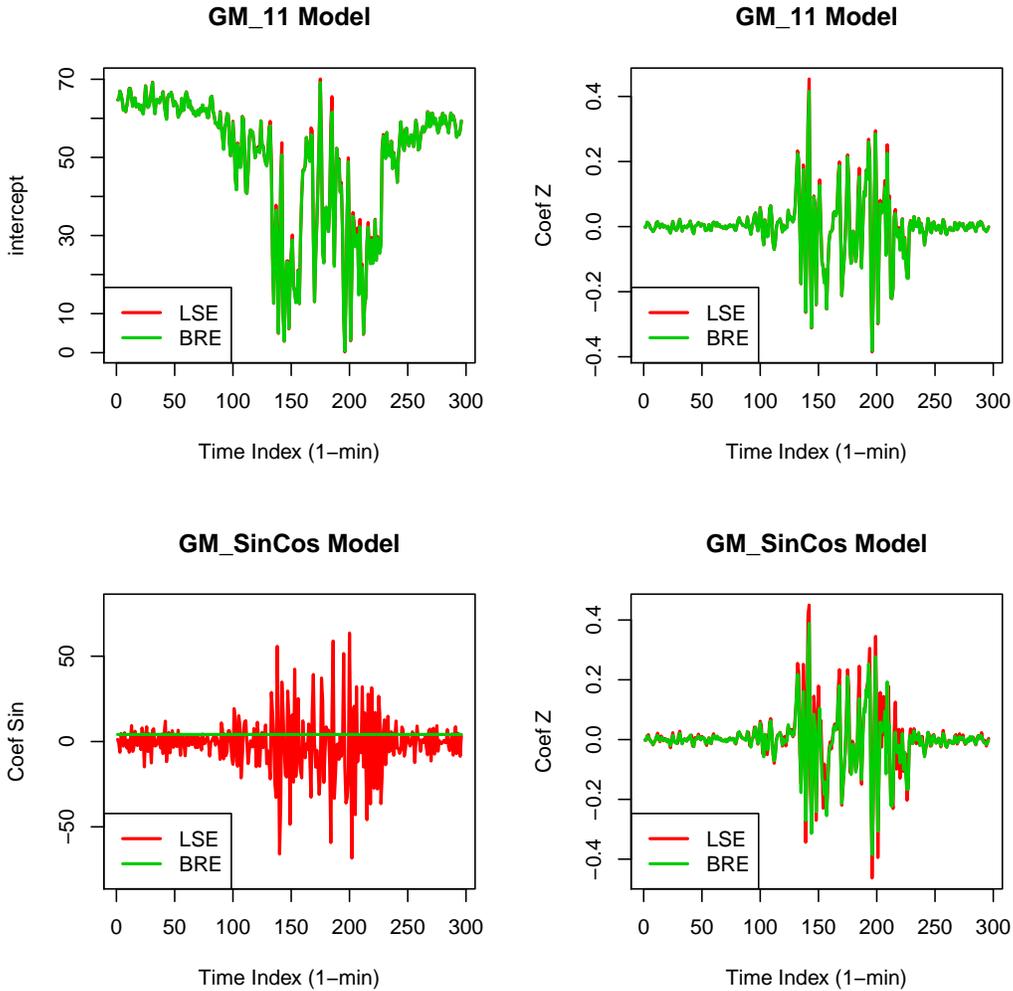}
            \caption{\textcolor{blue}{Example of evolution of parameter estimations LSE and BRE on February 19 (2)}}
            \label{fig_fqp2}
            \end{figure}
              
In Bayesian computations, we assigned noninformative priors to the regression parameters for 5000 samples and 500 burn in. Summary of the 1-step prediction results in mean squared errors 
 ($MSE$ = $\frac{\sum_{k=1}^{N}{\left(\hat{x}_(k)-x_(k)\right)^2}}{N}$) are given in Table~\ref{tab_accuracy} for two data series (February 18-1,February 19-2). Percent improvements in MSEs with Bayesian regression are also given in rows 3 and 4 (also given in Fig.~\ref{fig_imp}). We can see that GM\_11 was almost identical, but, other models are improved up to $45 \%$ with Bayesian regression estimation ($BRE$) compared to models with least squared estimations ($LSE$).
   	\begin{figure}[!h]
       \centering
       \includegraphics[width=0.8\linewidth]{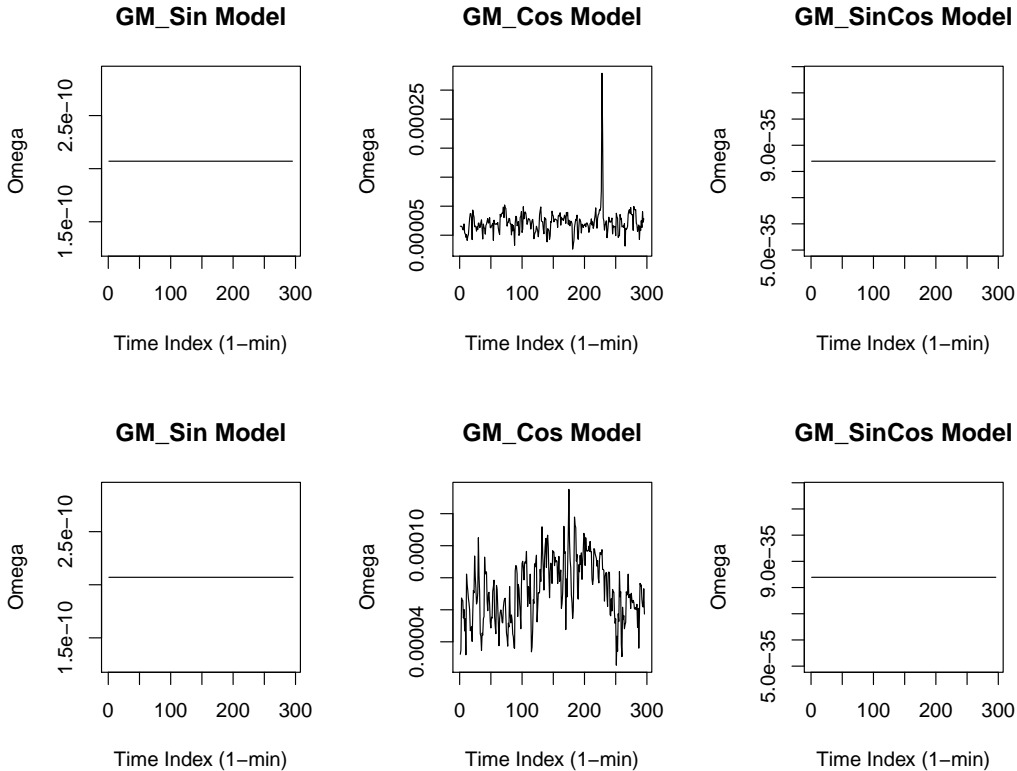}
       \caption{\textcolor{blue}{Bayesian regression estimation of $\omega$s in trigonometric Grey models for both days }}
       \label{fig_fqo}
       \end{figure} 
          
The results in Figures~\ref{fig_fq1}-\ref{fig_fqo} show the differences. The results for the more promising GVM and GM\_Cos models are shown in Fig.\ref{fig_fq1}. The results in these models are very close. For normal speed regimes, trigonometric models are performing better as GVM slightly overpredicts. When change occurs GVM predicts better. We see overprediction at changepoints with the GM$\_$Cos model.
               
The behavior of minute-by-minute parameter estimations is shown in Fig.\ref{fig_fqp1} and Fig.\ref{fig_fqp2}. We can see slight differences in estimations but the coefficients of trigonometric terms are different. Note that for the GM$\_$Cos model  with least squares estimation, $\omega$ was kept constant at $4.30$ and $9.30$ and is found by grid search. In the Bayesian setting, the behavior of Bayesian regression estimated $\omega$s in the trigonometric models are shown in Fig.~\ref{fig_fqo}. 
\section{Conclusions} 
In this study, we showed the efficacy of an online (real-time) Bayesian framework (utilizing WinBUGS in R) for GM(1,1), GM Verhulst, and three GM-trigonometric models using classical volatile time series traffic speed data. Based on the data used, we can see that using Bayesian regression for estimating GM parameters is promising. It can allow us to estimate the parameter $\omega$ in all the Grey Trigonometric models without going through the hassle of trying to find a value of $\omega$ by grid search.
However, we need to:
\begin{enumerate}[i.]
\item Test with more data series.This will involve more scenarios longer or shorter normal intervals.
\item Integrate Bayesian estimation as a function in $R$ to shorten run times. Currently, WinBUGS is called, opened, and closed. This process takes a few seconds for each rolling window.
\item As can be seen in (Fig.~\ref{fig_fqo}), $\omega$ in GM$\_$Sin and GM$\_$SinCos models is not changing like 
$\omega$  in   GM$\_$Cos model .  We need to focus on estimating $\omega$ in a better way.    
\end{enumerate}
\section*{Acknowledgments}
This research was partially supported by the Center for Connected Multimodal Mobility ($C^{2}M^{2}$) (USDOT Tier 1 University Transportation Center) headquartered at Clemson University, Clemson, South Carolina. It is also partially supported by U.S. Department of Homeland Security Summer Research Team Program Follow-On grant, NSF Grants Nos. 1719501, 1436222, 1954532, 1400991, EPSCoR Made in SC GEAR-CRP, U.S. Department of Education MSEIP Grant Award P120A190061, ​and NASA ULI project headquartered at University of South Carolina, Columbia. Any opinions, findings, and conclusions or recommendations expressed in this material are those of the authors and do not necessarily reflect the views of the Center for Connected Multimodal Mobility ($C^{2}M^{2}$) and the official policy or position of the USDOT/OST-R, or any State or other entity, and the U.S. Government assumes no liability for the contents or use thereof. 
\section*{Appendix}      
\begin{appendix}
    \lstset{caption={\textcolor{blue}{Bayesian Regression Model Fit for GM$\_$Cos model in WinBUGS within one-step speed prediction}}}
    \lstset{label={lst:alg1}}
     \begin{lstlisting}[style = R]

model{
  for (i in 1:N) {
  Y[i] ~ dnorm(mu[i],tau)
  mu[i] <- alpha + beta1 * x1[i]+beta2*cos(omega*x2[i])
  }
alpha ~ dnorm(0, 0.0001)
beta1  ~ dnorm(0, 0.0001)
beta2  ~ dnorm(0, 0.0001)
omega  ~ dchisqr(0.001)
tau ~ dgamma(0.001, 0.001)
sigma <- 1.0/sqrt(tau)
}
#Init and Data: x1,x2 are defined in R
\end{lstlisting}

\end{appendix}

\bibliographystyle{elsarticle-num}
\bibliography{bayesian}

\end{document}